\documentclass[sigconf]{acmart}

\usepackage{booktabs} 
\usepackage{float}
\usepackage{multirow}
\usepackage{lipsum}
\usepackage{graphicx}
\usepackage{csquotes}
\usepackage{array}
\newcolumntype{P}[1]{>{\centering\arraybackslash}p{#1}}
\newcolumntype{M}[1]{>{\centering\arraybackslash}m{#1}}

\setcopyright{rightsretained}

\acmDOI{10.475/123_4}

\acmISBN{123-4567-24-567/08/06}

\acmConference[CSCW'18]{Workshop of The 21st ACM Conference on Computer-Supported Cooperative Work and Social Computing}{November}{Jersey City, NJ}
\acmYear{2018}
\copyrightyear{2018}

\acmArticle{4}
\acmPrice{15.00}

\begin{document}
\title{Towards Connecting Experiences during Collocated Events through Data Mining and Visualization}


\author{Shuo Niu}
\orcid{Hide}
\affiliation{%
  \institution{Virginia Tech}
  \streetaddress{street address}
  \city{Blacksburg}
  \state{VA}
  \postcode{24060}
}
\email{shuoniu@vt.edu}

\author{Scott McCrickard}
\orcid{Hide}
\affiliation{%
  \institution{Virginia Tech}
  \streetaddress{street address}
  \city{Blacksburg}
  \state{VA}
  \postcode{24060}
}
\email{mccricks@cs.vt.edu}

\author{Steve Harrison}
\orcid{Hide}
\affiliation{%
  \institution{Virginia Tech}
  \streetaddress{street address}
  \city{Blacksburg}
  \state{VA}
  \postcode{24060}
}
\email{srh@vt.edu}
\begin{abstract}
Themed collocated events, such as conferences, workshops, and seminars, invite people with related life experiences to connect with each other. In this era when people record lives through the Internet, individual experiences exist in different forms of digital contents. People share digital life records during collocated events, such as sharing blogs they wrote, Twitter posts they forwarded, and books they have read. However, connecting experiences during collocated events is challenging. Not only one is blind to the large contents of others, identifying related experiential items depends on how well experiences are retrieved. The collection of personal contents from all participants forms a valuable group repository, from which connections between experiences can be mined. Visualizing same or related experiences inspire conversations and support social exchange. Common topics in group content also helps participants generate new perspectives about the collocated group. Advances in machine learning and data visualization provide automated approaches to process large data and enable interactions with data repositories. This position paper promotes the idea of \textit{event mining}: how to utilize state-of-the-art data processing and visualization techniques to design event mining systems for connecting experiences during collocated activities. We discuss empirical and constructive problems in this design space, and our preliminary study of deploying a tabletop-based system, \textit{BlogCloud}, which supports experience re-visitation and exchange with machine-learning and data visualization.
\end{abstract}

%
%



\maketitle

\section{Introduction}
Events such as conferences, workshops, or occasionally organized group talk and themed party, form a collocated social context for attendees to exchange work and life experiences \cite{Fischer2016CollocatedResearch}. People bring experiences to the event, sharing their own and listening to other's. Digital personal archives log life materials and support re-visiting and reminiscing life experiences \cite{Odom:2014:DSA:2611528.2557178,Sellen:2007:LTS:1240624.1240636}. Experiential items mentioned in conversations during the event are also documented in blogs, Twitter/Facebook posts, books, and other forms of life content. Experiences from all attendees form a collection of \textit{group content}, from which people learn information and understand the community. Social activities are commonly around life contents, such as sharing a blog collection about a trip, commenting social media topics one was following, or sharing an article one has read. Communicating about these contents depends on how well a speaker can recognize related experiences, and the degree the listener can capture other's experiences. However, even one person's experiential record could be huge. Exchanging and connecting these life contents is challenging without sufficient approaches to retrieve and connect life contents. 
\par
Prior studies have looked into system designs to re-present the records of life experiences. Interaction designs supporting re-visitation of personal life-logs seek to regenerate prior experiences by encouraging interaction with past photographs \cite{Odom:2014:DSA:2611528.2557178}, videos \cite{Kalnikaite:2010:LMS:1753326.1753638,Sellen:2007:LTS:1240624.1240636}, and geo-locations \cite{Thudt2016VisualData}. However, personal textual content such as blogs, articles, and social media posts have long been under-utilized to encourage re-visitation and support social exchange during collocated events. Personal blog archives take time to read, especially when they grows massively large but minimally organized \cite{Baumer:2008:ERR:1357054.1357228,Indratmo:2008:EBA:1385569.1385578}. Recent advancement in data science and visualization open new pathways to utilize them to augment social communication. Experience records contributed by participants form a group content repository, from which experience connections can be identified. Topics, common items, and related experiences can be mined from the large group content and visualized to support communication. The pattern of how experiences are related brings up new perspectives about the participants as an entity. Supporting \textit{event mining} for hybrid co-located events needs to consider ways to construct group content repository, interactions with the group content, and knowledge from group content for reflection. This position paper explores the design space of mining the life contents of collocated event participants to support connecting experiences. We first discusses empirical and constructive problems in this design space, and then demonstrate our initial exploration of an interactive blog visualization, BlogCloud, which incorporates machine learning and visualization techniques to support re-visitation and reflection of large personal blogs. 
\section{Background}
Harrison et al. \cite{Harrison1996Re-Place-ingSystems} defined the role of \textit{space} and \textit{place} in CSCW design as \textit{``space is the opportunity; place is the understood reality"}, and CSCW systems should support ``place-making" in physical and virtual space. Collocated events forms an opportunity space for participants to meet people with similar experiences. The social activities during the event make it an understood place to seek such experiences. A hybrid event not only brings people physically together, but also provide digital channels to bridge participants and enhance connection. Digital and virtual channels of communication construct novel attending experiences and inspire new perspectives about group. To make a collocated event a better place to connect similar experiences, CSCW systems should not only support communications in the current space, but also consider experience records from participant's life and connecting different experiences.
\par
People spend time crafting textual posts, recording marks in life, and capturing essences of thoughts \cite{Nardi:2004:WWB:1035134.1035163}. However, years after crafted, the large volume of minimally organized personal content could hinder the re-visitation and reflection. Re-collecting and retrieving information about things in personal experience costs time, which is difficult without text-processing techniques. Looking up and collecting information about particular items rely on memory, and cost time with big content volume \cite{Nardi:2004:BSA:1031607.1031643}. In co-located space, connecting personal content from different people needs automated approaches to identify related experiential items and common themes and topics. To support connecting experiences during co-located activities, machine learning (ML) and visualization (VIS) provide new opportunities to support collaborative re-visitation to the experience records \cite{Dove:2017:UDI:3025453.3025739}. Related things and similar experiences can be algorithmically mined from the time stream and reordered in a visualization. ML and VIS could build new perspectives to encourage reflection on each other's experiences, or explore integrated group content as a whole. Devices such as mobile phones, large interactive surfaces, VR/AR gadgets can be used to enable hybrid interactions with the each other's experiences, and trigger social communication about the group content.
\section{Problem Space}
Towards better CSCW systems for hybrid events, we consider the design space of \textit{event mining}, which incorporates data mining and visualization to support connecting experiences between collocated participants. Based on the timeline of an event, we consider \textit{preparation}, \textit{interaction}, and \textit{reflection} as three stages to implement event mining systems. Preparation stage gathers life content from individual attendees and form a focused repository. In interaction stage meaningful experience connections are recognized from group content and visualized for collaborative interaction. Reflection stage concerns how to archive the group content and people's interaction during and after the events.
\begin{figure*}[h!]
\includegraphics [width=\textwidth]{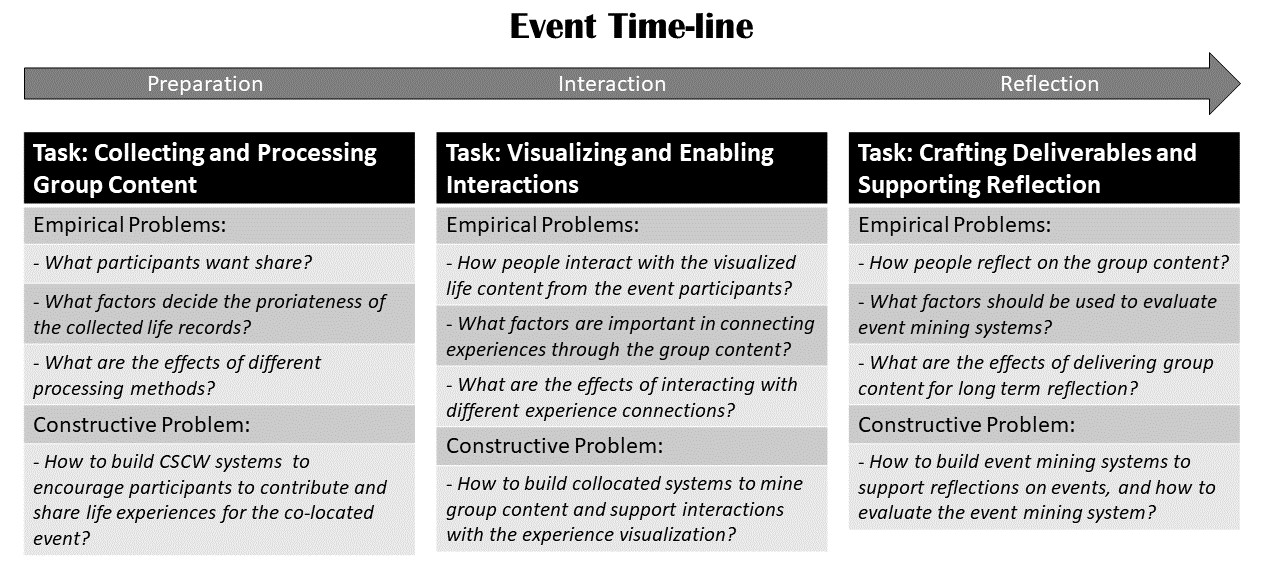}
\caption{Empirical and constructive problems of event mining}
\label{fig:all_viz}
\end{figure*}
\subsection{Preparation}
Mining the experience records of collocated participants needs to collect personal contents from people. Depend on the theme of the event, different types of documents can be collected. Life contents could be articles about personal life, such as blogs, Twitter/Facebook posts, and books people wrote. For work events such as conferences and seminars, contents such as publications, job descriptions, and CV/resumes can also be sources of repository. Event mining during preparation stage should encourage people to make contributions to the event. Potential ways to collect data from participants are either utilizing life contents people already posted on social media, or asking participants to suggest things to be shared with others. Either ways involve questions about what kind of data are appropriate to be connected, and how could the data be meaningfully processed. People might hesitate about sharing individual contents due to issues such privacy consideration, social anxiety, and individual roles in the events. Motivation factors need to be evaluated and understood to avoid privacy and social concerns. Approaches to collect personal contents also need to identify construct of the experience records. Factors such as formats of the data, amount of data, and time-span of collection need to be considered. Textual features need to be feasible and desirable to be used to connect experiences. This knowledge will benefit building better machine learning techniques to mine the group content.
\subsection{Interaction}
During the event, people visit the group content and connect to each other's experiences. This phase focuses on designing collaborative tools which present group content and support connecting experiences. Data visualizations could be used to present group content, but they need to consider manners to show individual content to the public. Visualizations of group content need to motivate people to share and reflect on each other's experiences. Collaborative interactions can be interpreted to reorganize the group content and retrieve experiences dynamically. Paradigms to present the connect experiences may lead to different social communication effects. For the constructive problem, practitioners need to consider forms of devices to support viewing event mining results. Multi-user multi-touch displays in public space support visualizing large data collection and enables people to interact at the same time \cite{Peltonen2008ItsCentre, Jacucci2010WorldsDisplay, Niu2017AnCollaboration}. This type of devices support connecting experiences between dyads or trios. Distributed devices such as mobile phones or computers allow more participants to access mining results on the internet, but this implementation needs to consider how to motivate social exchange between the collocated participants.
\subsection{Reflection}
Studying people's reflection on the group content need to understand the phenomenon of how people perceive and interpret the ML-processed content. Factors of event mining which raise awareness of experience connections and trigger conversation need to be identified and understood \cite{Niu2017InvestigatingDisplays}. Evaluation methods are needed to assess people's reflective activities and effectiveness of the data mining approaches. Summaries and deliverables of group content can benefit re-visiting and reminiscing the attending experience. After events, the methods to archive group content and event activities should be designed to support long term reflection. Event mining needs to record not only the group content, but also people's interactions during the events. Interactive components including marks, notes, and interaction data could be used to build novel reflective materials for hybrid events. The construction of event mining systems should consider how to design for short and long term reflection on the group content. Design opportunities exist in converting the event mining results to take-away items such as mementos and event archives.
\section{BLOGCLOUD: BLOG REFLECTION TOOL}
\begin{figure}
\includegraphics[width=0.43\textwidth]{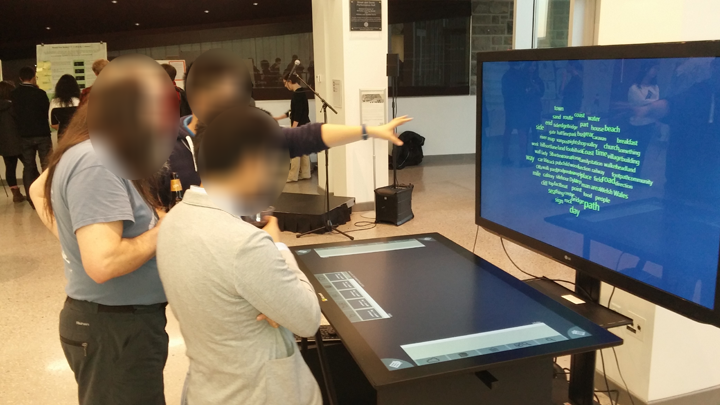}
\caption{The BlogCloud System. The tabletop supports blog searching and viewing. The vertical display shows a visualization.}
\label{fig:interface}
\end{figure}

In this section, we introduce \textit{BlogCloud}, an interactive blog reflection tool implemented with machine learning techniques and text visualization (Figure \ref{fig:interface}). BlogCloud is designed for the interaction stage, which supports participants come and interact with the experience records in an event setting. BlogCloud seeks to provide overviews of experiential items, support organization and reminiscence of experiences, and allow searching symbols from the temporally-ordered content. To support interaction in collocated space, BlogCloud is implemented with a dual-display design - a multi-touch tabletop and a vertical display. Though BlogCloud current only supports interacting with one participant's blog set, it is an preliminary exploration of how people interact with the ML-processed content and generate new perspectives about experiences.

\begin{figure}
\includegraphics[width=0.3\textwidth]{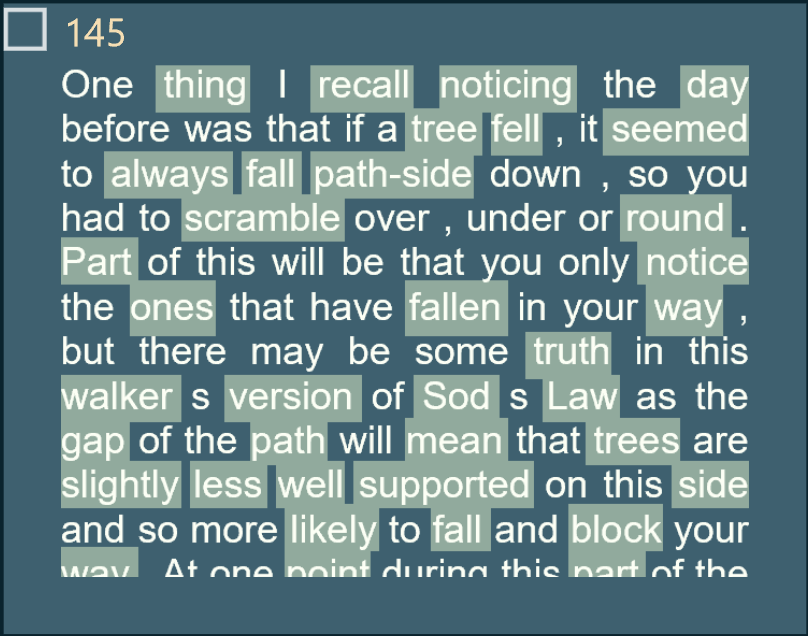}
\caption{A paragraph card. Words with background color are recognized keywords.}
\label{fig:blogcard}
\end{figure}
Experiential items in Blogs are recorded in reverse chronological order \cite{Nardi:2004:WWB:1035134.1035163}. BlogCloud incorporates natural language processing (NLP) techniques to break the flow of blogs, and to chuck and clean the blogs into elementary experiential items. A \textit{paragraph} is used as the basic reflection unit, since paragraphs usually focus on a few inter-related experiential items that are lexically self-contained and map well for machine learning. For each paragraph, nouns, verbs, adjectives and adverbs are marked as keywords with coreNLP (a NLP toolkit for part-of-speech recognition. Digital cards present he processed paragraphs on the multi-touch tabletop (Figure \ref{fig:blogcard}). The cards can be moved, rotated and zoomed with multi-finger gestures. Zooming a card shows the blog paragraph in multiple levels of detail. Segmentation and term-weighting techniques enable the system to process experiential items in blogs. 
\par
A word cloud is visualized on the vertical display for each paragraph group created by the user, with four visualization components: highlighted words, feature words, number of similar documents, and average sentiment values (Figure \ref{fig:interface}). The highlighted words are presented with yellow borders. The size of the words is proportional to the weight of words in the paragraph group. A count of similar paragraphs is presented in a circle. A strip with the number of associated blogs connects each cloud pair. The width of the strip is proportional to the number of blog paragraphs associated with both groups.
\begin{figure}
\includegraphics[width=0.47\textwidth]{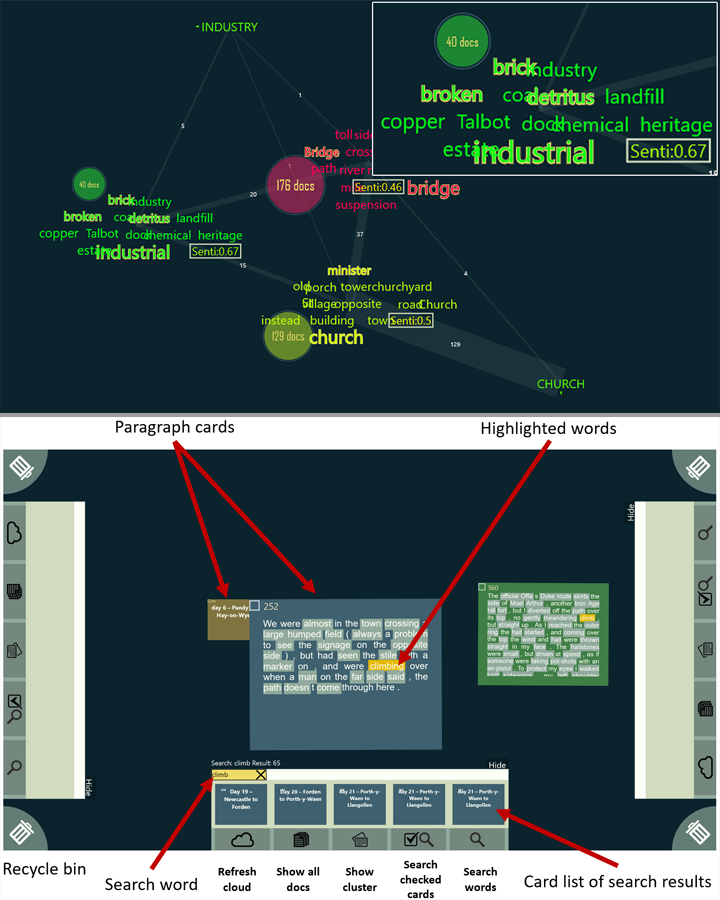}
\caption{Top: Touch interface runs on the tabletop display. Bottom: Connected word cloud visualizing paragraph groups and connections between groups and search keys. Up-right corner: Word cloud for one paragraph group.}
\label{fig:interface2}
\end{figure}
\subsection{How Blog Authors Reflect}
Two blog authors attended a workshop reception and used the tool. The third blog author was unable to attend the reception, so he used the system during a separate lab visit. The workshop has a hiking theme, and attendees are either lovers of outdoor activities or experts studying the trail. In the author session, each blog author spent 20-40 minutes using BlogCloud to explore his or her own hiking blogs. During the workshop, other attendees occasionally joined the conversation and engaged with the system. However, the blog author remained the primary user of the system throughout the session. 
\subsubsection{Reminisce about particular experiences}
During the author session, blog authors reminisced particular experiences and shared with other people around. When A1 searched a word ``climb", he noticed a paragraph about a gradually narrowed cliff path. A1 reminisced the experience on the cliff path by telling a story. \textit{``It was a path on the cliff. The first time when I passed it, it was this wide,"} A1 opened his arms to show how wide it was. \textit{``But the second time when I was there it was barely my feet's width. I have to pass it with my body clinging to the cliff and be careful."} The particular experience of passing a terrifying cliff path re-presented this kind of ``climb" experience. A2 noted reflecting on the word cloud surfaced her memory about her experiences. She commented about the system \textit{``the connections I got to explore and the way that the semantic connections surfaced specific memories... it was a really cool experience to see words connected that then acted as triggers to surface memories that I otherwise would not necessarily have been thinking about. The connections were the interesting part, much more than just reflecting on some experience alone."} Reading the visualization raised the reflection on particular experiences, which reflects that re-presentation of a general experience turns into a particular one which represent that kind of experience.
\par
\subsubsection{Making sense of symbols}
\begin{figure}[!ht]
\includegraphics[width=0.47\textwidth]{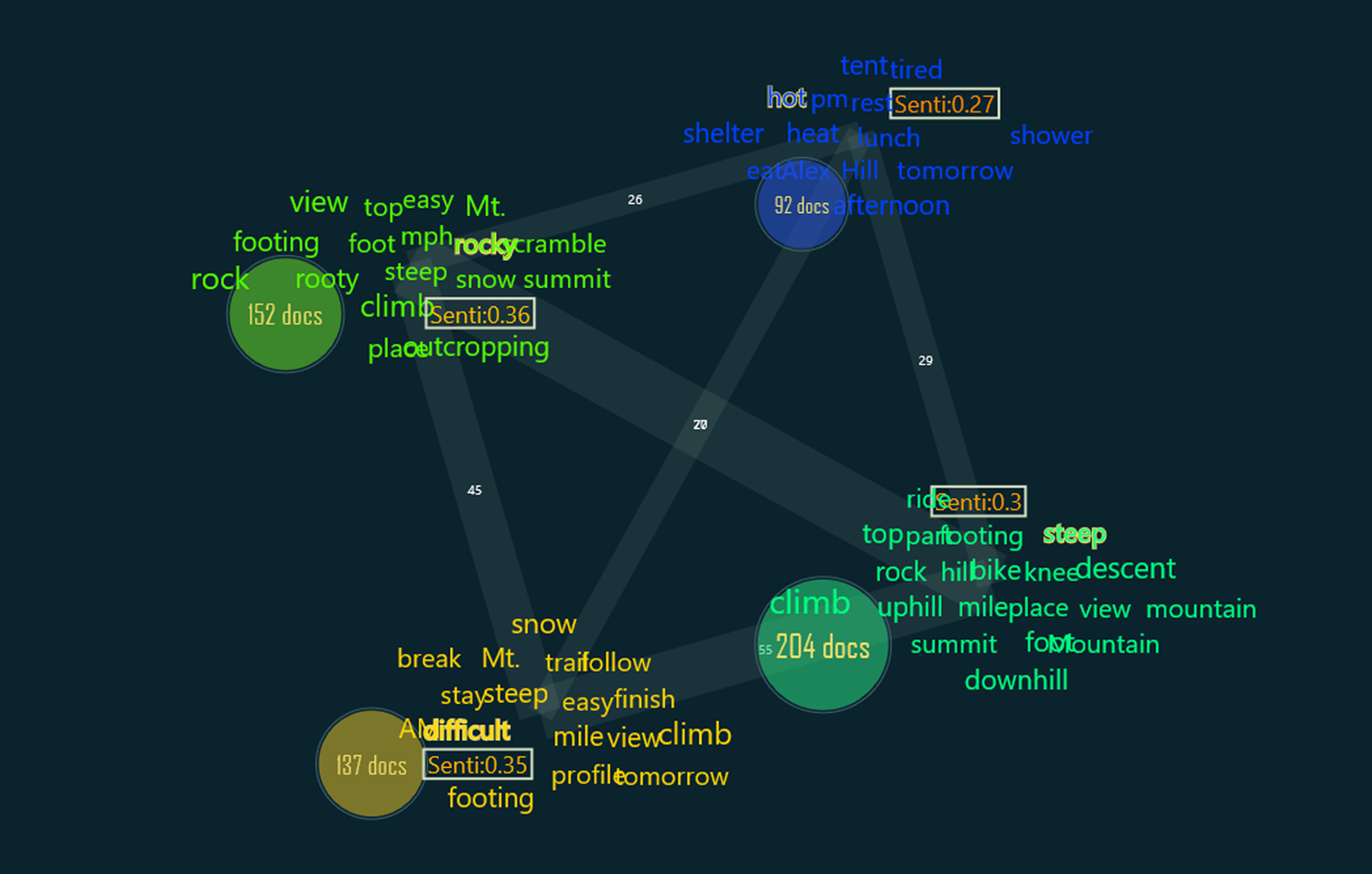}
\caption{Four word clouds created by A3.}
\label{fig:A3wordcloud}
\end{figure}

BlogCloud offers a different view of symbols for reflection. In the author session, blog authors had sense-making activities with their own blogs. The visualization of paragraph groups was used by A3 to make sense of symbols, as a way to reflect on his own blog from a different perspective. A3 used the visualization to compare recurring experiences. He searched ``difficult", ``rocky", ``steep", and ``hot", creating four corresponding pragraph groups. After reading the connecting strips (Figure \ref{fig:A3wordcloud}), A3 commented, \textit{``it makes sense to me since 'steep' is more related to 'rocky' and 'difficult' than 'hot'"}. When seeing ``bike" had more documents than ``scooter", he commented \textit{``[the visualization] reflects that I used bike} more than \textit{scooter for transportation"}. When speculating that ``cold" was mentioned more than ``hot" in his blogs, he paused for a while and said, \textit{``'cold' is bigger than 'hot', hmm... maybe that is because it took me more time hiking the colder north part, than the warmer south."} Though A1 did not specifically use BlogCloud to compare search results, he found sense making important for reflection on personal content. He commented, \textit{``Somebody else did a grounded theory analysis of my blogs. They surprised me as they (themes of blogs) were all about social experience whereas I said that the vast majority of the time I was alone. Although you have written the blog, it does not mean you fully grasp the big themes that run through it ... hence the need for sense-making tools."}
\subsubsection{Recovering lost memory}
Experience documentation has been recognized as the most prevalent purpose for blogging \cite{Nardi:2004:WWB:1035134.1035163}. However, considering the large number of the blog posts and time passed since creation, less significant experiential items might be lost in memory, requiring effort to be re-generated. We notice all three blog authors found words in the word cloud or paragraphs on the cards that they did not recognize. But blog authors did not skip these clues; they would collect more information to recover their lost memory. When the visualization showed ``dead" and ``porpoise" as keywords in one paragraph group. One attendee asked whether seeing ``dead porpoise" made A1 sad? A1 could not remember where he wrote about porpoise, and searched ``porpoise" and learned that he once thought the cracked wood looked like ``snout of a dead porpoise". A3 searched words Jersey and several people's names came out. He searched a name among the ``Jersey" group and found the paragraph with those names. Through reading the paragraph he recalled the moment he met other hikers in New Jersey. 
\section{Conclusions and Future Directions}
CSCW community meets the era when data mining and artificial intelligence are incorporated in HCI and CSCW applications \cite{Dove:2017:UDI:3025453.3025739}. Connections during hybrid events are not only through the channel of face-to-face conversation, but could be ``mined" from participants' life records. This position paper explores \textit{event mining} on group content - using data mining techniques to process the life contents collected from event participants and visualized for connecting experiences. We explore the problem space and design space for event mining systems, and share our experience deploying a ML- and VIS- augmented blog exploration tool during a themed workshop.
\par
Moving forward, we will explore designs and applications which incorporate different types of life records to support interactive exploration of experiences. Design options to trigger conversations and support social exchanges will be compared to contribute new knowledge to support hybrid events. Meaningful reflection activities such as reminiscence, sense-making and memory recovery will be further examined with different group content and group content representations. Following the empirical and constructive problem structure, we seek to gain a deeper understanding of design opportunities and challenges in utilizing data mining techniques to support hybrid events.
\bibliographystyle{ACM-Reference-Format}
\bibliography{my.bib}

\end{document}